\documentclass[a4paper,11pt]{article}
\usepackage{epsfig}
\usepackage{axodraw}
\usepackage{graphicx}

\newcommand{\la}{{\lambda}}

\newcommand{\be}{\begin{equation}}
\newcommand{\ee}{\end{equation}}
\newcommand{\beq}{\begin{equation}}
\newcommand{\eeq}{\end{equation}}
\newcommand{\bea}{\begin{eqnarray}}
\newcommand{\eea}{\end{eqnarray}}
\newcommand{\br}{\begin{eqnarray}}
\newcommand{\er}{\end{eqnarray}}
\newcommand{\ba}{\begin{array}}
\newcommand{\ea}{\end{array}}
\newcommand{\bi}{\begin{itemize}}
\newcommand{\ei}{\end{itemize}}
\newcommand{\bn}{\begin{enumerate}}
\newcommand{\en}{\end{enumerate}}
\newcommand{\bc}{\begin{center}}
\newcommand{\ec}{\end{center}}

\def\bY{{\bf Y}}
\def\bA{{\bf A}}

\def\bU{{\bf U}}

\def\bm{{\bf m}}

\def\tl{{\tilde{L}}}

\def\te{{\tilde{e^c}}}

\def\td{{\tilde{d^c}}}
\def\tq{{\tilde{Q}}}
\def\tu{{\tilde{u^c}}}

\def\unity{{\hbox{1\kern-.8mm l}}}
\newcommand{\ov}{\overline}

\newcommand{\no}{\nonumber}

\newcommand{\ga}{\gamma}

\newcommand{\gsim}{\lower.7ex\hbox{$\;\stackrel{\textstyle>}{\sim}\;$}}
\newcommand{\lsim}{\lower.7ex\hbox{$\;\stackrel{\textstyle<}{\sim}\;$}}

\begin{document}
\begin{flushright}
{DFPD-06/TH/24}
\end{flushright}
\vspace{0.5cm}

\begin{center}
{\huge \bf \sf \mbox{ Gauge and Yukawa mediated supersymmetry
breaking} in the triplet seesaw
scenario}\\[1cm]

{ {\large\bf Filipe R. Joaquim$^{a,b,\ast}$ and Anna Rossi$^{a,\S}$}
}
\\[7mm]
{\it $^a$ Dipartimento di Fisica ``G.~Galilei'', Universit\`a di
Padova
I-35131 Padua, Italy}\\[3mm]
{\it $^b$ Istituto Nazionale di Fisica Nucleare (INFN), Sezione di
Padova, I-35131 Padua, Italy}
\\[1cm]
\vspace{-0.3cm}

{\tt  $^\ast$\,E-mail: \hspace*{-0.3cm}joaquim@pd.infn.it\\
$\!\!\!^\S$\,E-mail: \hspace*{-0.3cm}arossi@pd.infn.it}

\vspace{1cm}

{\large\bf ABSTRACT}
\renewcommand{\baselinestretch}{1.5}
\end{center}

\begin{quote}
{\large\noindent We propose a novel supersymmetric unified scenario
of the triplet seesaw mechanism where the exchange of the heavy
triplets generates both neutrino masses and soft supersymmetry
breaking terms. Our framework is very predictive since it relates
neutrino mass parameters, lepton flavour violation in the slepton
sector, sparticle and Higgs spectra and electroweak symmetry
breakdown. The phenomenological viability and experimental
signatures in lepton flavor violating processes are discussed.}
\end{quote}

\renewcommand{\baselinestretch}{1.5}
\vspace*{0.5cm} \large Modern particle physics has been confronting
the intriguing issue of neutrino mass generation and its
phenomenological implications. From the theoretical point of view,
the well-celebrated seesaw mechanism provides a natural explanation
for the generation of  neutrino masses and their suppression with
respect to the other fermion masses of the Standard Model (SM). In
its most popular versions, the seesaw mechanism is realized either
by exchanging singlet fermions $N$ (type I) ~\cite{seesaw}, or a
$SU(2)_W$ scalar triplet $T$ with non-zero hypercharge (type
II)~\cite{ss2}, at a high scale $M_L$. An attractive feature of the
supersymmetric version of the above scenarios is that lepton flavor
violating (LFV) processes (otherwise unobservable) can be enhanced
through  one-loop exchange of lepton superpartners if their masses
do not conserve flavor. Regarding this aspect, most of the
literature has been focussing on the most conservative scenario of
universal sfermion masses at high energy, as in minimal supergravity
or gauge mediated supersymmetry (SUSY) breaking models. In such
cases, flavor non-conservation in the sfermion masses arises from
renormalization group (RG) effects due to flavor-violating Yukawa
couplings ~\cite{lfv,rgess,ar}. We recall that in the triplet seesaw
the flavor structure of the slepton mass matrix $\bm^2_\tl$ after RG
running can be univocally determined in terms of the low-energy
neutrino parameters ~\cite{ar}. This is in contrast with the type-I
seesaw where the structure of $\bm^2_\tl$ cannot be unambiguously
related to the neutrino parameters.

In this Letter we present a novel supersymmetric scenario of the
triplet seesaw mechanism in which the soft SUSY breaking (SSB)
parameters in the minimal supersymmetric extension of the SM (MSSM)
are generated at the decoupling of the heavy triplets and the mass
scale of such  SSB terms is fixed {\it only} by the triplet SSB
bilinear term $B_T$. This scenario is highly predictive since it
relates neutrino masses, LFV in the sfermion sector, sparticle and
Higgs spectra and electroweak symmetry breaking (EWSB).

The supersymmetric version of the type-II seesaw requires
introducing the triplets as super-multiplets $T, \bar{T}$ in a
vector-like $SU(2)_W\times U(1)_Y$ representation, $T\sim (3,1),
\bar{T} \sim (3,-1)$. In order to preserve successful gauge coupling
unification, we embed our framework in a $SU(5)$ grand unified
theory (GUT) \cite{ar} where the triplet states fit into the 15
representation $15 = S + T +Z$ transforming as $ S\sim
(6,1,-\frac23), ~T \sim (1,3,1),~ Z\sim (3,2,\frac16)$ under $SU(3)
\times SU(2)_W \times U(1)_Y$ (the $\ov{15}$ decomposition is
obvious).
The SUSY breaking mechanism is parametrized by a gauge singlet
chiral supermultiplet $X$, whose scalar $S_X$ and auxiliary $F_X$
components are assumed to acquire a vacuum expectation value
 through some unspecified dynamics in the secluded sector. It
is suggestive for our discussion to assume that the  $SU(5)$ model
conserves the combination $B-L$ of baryon and lepton number. As a
result, the relevant superpotential reads
\bea
&&\!\!\!\!\!\!\!\!\!\!W_{SU(5)}=\! \frac{1}{\sqrt2}(\bY_{15} \bar5~
15 ~\bar5 +
 \la {5}_H ~\ov{15}~ {5}_H)
 + \bY_5 \bar5 ~ \bar5_H 10 \no \\
&&\!\!\!\!\!\!\!\!\!\phantom {W_{SU(5)}=} + \bY_{10} 10 ~10 ~5_H +
M_5  5_H~\bar5_H  + \xi X 15 ~ \ov{15} , \label{su5}
\eea
where  the matter multiplets are understood as $\bar5  = (d^c, L)$,
$10 = (u^c,e^c,Q)$ and the Higgs doublets fit with their coloured
partners, $t, \bar{t}$ like ${5}_H = (t, H_2), \bar{5}_H = (\bar{t},
H_1)$. The $B-L$ quantum numbers are a combination of the
hypercharges and the following charges: $Q_{10}= \frac15,\,
Q_{\bar{5}} = - \frac35, \, Q_{{5}_H} =-\frac25, \,Q_{\bar{5}_H }=
\frac25,\, Q_{15} = \frac65, \,Q_{\ov{15}}= \frac45$ and $Q_{X} =
-2$. The form of $W_{SU(5)}$ implies that  the $15, \ov{15}$ states
play the role of {\it messengers} of both $B-L$ and SUSY breaking to
the visible (MSSM) sector thanks to the coupling with $X$. Namely,
while $\langle S_X\rangle$ only breaks $B-L$, $\langle F_X\rangle$
breaks both SUSY and $B-L$. These effects are  parametrized by the
superpotential mass term $M_{15} 15~\ov{15}$, where $M_{15} = \xi
\langle S_X\rangle$, and the bilinear SSB term $-B
M_{15}15~\ov{15}$, with $ B M_{15}= - \xi \langle F_X\rangle$. Once
$SU(5)$ is broken to the SM group  we find~\cite{ar}, below the GUT
scale $M_G$,
 \bea
&&\!\!\!\!\!\!\!\!\!\!\!\!\!\!W = W_0 + W_T + W_{S,Z} \no \\
&&\!\!\!\! \!\!\!\!\!\!\!\!\!\!W_0 =  \bY_e  e^c H_1  L
+\bY_d d^c H_1  Q + \bY_u  u^c Q  H_2  + \mu H_2 H_1 \no \\
&& \!\!\!\! \!\!\!\!\!\!\!\!\!\!\!W_T= \frac{1}{\sqrt{2}}(\bY_{T} L
T L  + \la H_2 \bar{T} H_2) +
  M_T T \bar{T} \label{WT} \no \\
&&\!\!\!\! \!\!\!\!\!\!\!\!\!\!\!\! W_{S,Z}= \frac{1}{\sqrt{2}}\bY_S
d^c S d^c + \bY_Z  d^c  Z L + M_Z Z\bar{Z}+M_S S\bar{S}
.\label{su5b} \eea%
Here, $W_0$ denotes  the MSSM superpotential,  $W_T$ contains the
triplet Yukawa and mass terms, and $W_{S,Z}$ includes the couplings
and masses of the colored fragments $S,Z$. As in \cite{ar}, we have
relaxed the strict $SU(5)$ symmetry relations for the Yukawa
interactions and mass terms by allowing $SU(5)$ breaking effects,
induced, for example, by  adjoint $24$-insertions, such as $\bY_5 =
\bY^{(0)}_5 + \bY^{(1)}_5 24/\Lambda +  \ldots$ with a cut-off scale
$\Lambda> M_G$. These insertions are necessary to correct the
relation $\bY_e= \bY^T_d$ and to solve the doublet-triplet splitting
problem. For the sake of simplicity, we take $M_T = M_S = M_Z$ and
$\bY_{S},\bY_{Z}\ll \bY_T$ at $M_{G}$ (possibly due to
24-insertions), which does not alter the major point of our
discussion. The $SU(5)$ scenario with $\bY_{S}=\bY_{Z}= \bY_T$
implies correlations between LFV and quark flavour violation; this
case will be considered in detail  in \cite{af2}.
In eq.~(\ref{su5b}), $W_T$ is responsible for  the realization of
the seesaw mechanism. Actually, at the scale $M_T$ the triplets act
as tree-level {\it messengers} of lepton number and flavor
violation\footnote{Beneath the scale $M_G$, baryon number is
conserved since the colored partners $t, \bar{t}$ are understood to
be decoupled.} via the symmetric Yukawa matrix $\bY_T$, generating
the $d=5$ effective operator $\la \bY_T (LH_2)^2/M_T$. Subsequently,
at the electroweak scale the Majorana neutrino mass matrix is
obtained
 \be \label{T-mass}
{\bf m}^{ij}_\nu =\frac{ \la\langle H_2\rangle^2}{M_T} \bY^{i j}_T ,
~~~ i,j=e,\mu,\tau . \ee
In the basis where $\bY_e$ is diagonal, it is apparent that all LFV
is encoded in  $\bY_T$. Namely, the nine independent parameters
contained in $\bm_\nu$  are directly linked to the neutrino
parameters according to $\bm_\nu = \bU^* \bm^D_\nu \bU^\dagger$,
where $\bm^D_\nu = {\rm diag}(m_1,m_2, m_3)$ are the mass
eigenvalues, and $\bU$ is the leptonic mixing matrix.

Regarding the SSB term one has, in the broken phase,
 \be \label{ssb}
-B_T M_T (T\bar{T} + S\bar{S} + Z\bar{Z}) + {\rm h.c.},
\ee%
where $B_T \equiv B_{15}$. These terms lift the tree-level mass
degeneracy in the MSSM supermultiplets. Indeed, at the scale $M_T$,
all the states $T, \bar{T}, S, \bar{S}$ and $Z, \bar{Z}$ are {\it
messengers} of SUSY breaking to the MSSM sector via gauge
interactions, as it happens in conventional gauge-mediation
scenarios \cite{GR-PR}. However, in our framework the states $T,
\bar{T}$ also transmit SUSY-breaking  via Yukawa interactions.
Finite contributions  for the trilinear couplings of the
superpartners with the Higgs doublets, $\bA_e, \bA_u, \bA_d$, the
gaugino masses $M_a~ (a=1,2,3)$ and the Higgs bilinear term $- B_H
\mu H_2 H_1$ emerge at the one-loop level:%
\bea
 {\bA}_e & = &
\frac{3 B_T}{16 \pi^2} \bY_e \bY^\dagger_T \bY_T\,,\nonumber \\
 {\bA}_u &= &
\frac{3 B_T}{16 \pi^2} \bY_u  |\la|^2\,, \nonumber \\
 {\bA}_d &=& 0\,, \nonumber \\
 M_a &= &  \frac{7 B_T }{16 \pi^2}\,g_a^2 \,,\nonumber \\
B_H &=& \frac{3 B_T}{16 \pi^2}  |\la|^2~,
 \label{finite}
\eea ($g_a$ are the gauge coupling constants).
As for the SSB squared scalar masses, the leading ${\cal
O}(F_X^2/M^2_T)={\cal O}(B^2_T)$ contributions do not emerge at
one-loop level~\cite{dns}, but instead at two-loop\footnote{ Such
${\cal O}(F_X^2/M^2_T)$ two-loop contributions dominate over the
${\cal O}(F_X^4/M^6_T) ={\cal O}(B^4_T/M^2_T)$ one-loop ones for
$M_T > (4\pi Y_T/g^2) B_T$, which is  indeed fulfilled in our
analysis.}:
\bea \!\!\!\!\bm^2_{\tl}& =& \frac{|B_T|^2}{(16 \pi^2)^2} \left[
\frac{21}{10} g^4_1 + \frac{21}{2} g^4_2 - \left(\frac{27}{5} g^2_1
+21 g^2_2\right)\bY^\dagger_T\bY_T  + 3\bY^\dagger_T \bY^T_e \bY^*_e
\bY_T + 18  (\bY^\dagger_T\bY_T)^2 \right. \no \\
&& \!\!\!\!\!\!\!\! \left.+ 3\,{\rm Tr}(\bY^\dagger_T\bY_T)
\bY^\dagger_T\bY_T \right] \no \\
\!\! \bm^2_{\te}& =& \frac{|B_T|^2}{(16 \pi^2)^2} \left[\frac{42}{5}
g^4_1 - 6 \bY_e \bY^\dagger_T\bY_T\bY^\dagger_e \right]
\no \\
\!\!\bm^2_{\tq}& = &\frac{|B_T|^2}{(16 \pi^2)^2} \left[\frac{7}{30}
g^4_1 + \frac{21}{2} g^4_2 + \frac{56}{3} g^4_3
- 3  |\la|^2 \bY^\dagger_u \bY_u\right]\no \\
\!\!\bm^2_{\tu} &= &\frac{|B_T|^2}{(16 \pi^2)^2} \left[
\frac{56}{15} g^4_1 +\frac{56}{3} g^4_3
- 6  |\la|^2 \bY_u \bY^\dagger_u\right]\no \\
\!\!\bm^2_{\td}& = &\frac{|B_T|^2}{(16 \pi^2)^2} \left[\frac{14}{15}
g^4_1
 +\frac{56}{3} g^4_3 \right]\no \\
\!\! m^2_{H_1} &= &\frac{|B_T|^2}{(16 \pi^2)^2} \left[
 \frac{21}{10} g^4_1 +   \frac{21}{2}  g^4_2
\right] \no \\
\!\! m^2_{H_2}& = &\frac{|B_T|^2}{(16 \pi^2)^2} \left[ \frac{21}{10}
g^4_1 +   \frac{21}{2}g^4_2
 - \left(\frac{27}{5} g^2_1
+21 g^2_2\right)|\la|^2  + 9 |\la|^2 {\rm Tr}(\bY_u\bY^\dagger_u) +
21 |\la|^4 \right]. \label{soft2}
\eea
The results (\ref{finite}) and (\ref{soft2}) can be obtained either
by diagrammatic computations or from generalization of the wave
function renormalization method proposed in \cite{GR}.

Notice that the generation of all the SSB gaugino masses requires
the presence of the complete 15 representation. More specifically,
$M_1, M_2$ and $M_3$ arise from the exchange of the $(T, S, Z)$,
$(T, Z)$ and $(S, Z)$ states, respectively.
%
The expressions in eqs.~(\ref{finite}) and (\ref{soft2}) hold at the
decoupling scale $M_T$ and therefore are meant as boundary
conditions for the SSB parameters which then undergo (MSSM) RG
running  to the low-energy scale $\mu_{SUSY}$.
In particular, we observe that the  Yukawa couplings $\bY_T$ induce
LFV to $\bA_e$, to the scalar masses $\bm^2_\tl$ and to a much less
extent in $\bm^2_\te$. This feature makes the present scenario
different from pure gauge-mediated models \cite{GR-PR} where flavor
violation comes out naturally suppressed (for other examples of
Yukawa mediated SUSY breaking, see {\it e.g} \cite{dns,DS}). We
suppose that  possible gravity mediated contributions $\sim F
/M_{pl}$ (where
 $ F^2 =   \langle |F_X|^2\rangle +  \ldots $ is the sum of $F$-terms
in the secluded sector) are negligible. This is the case if $M_T \ll
10^{16}~{\rm GeV}~ \xi \langle F_X\rangle/F$. Furthermore, it is
necessary that $ \xi \langle F_X\rangle  < M_T^2$ (or $B_T < M_T$)
to avoid tachyonic scalar messengers.

It is worth stressing that here the LFV entries $(\bm^2_\tl)_{ij}\,
(i\neq j)$ show up as finite radiative contributions induced by
$B_T$ at $M_T$, and they are not essentially modified by the (MSSM)
RG evolution to low-energy. This is different from a previous work
\cite{ar} where a common SSB scalar mass $m_0 \sim {\cal O}(100~{\rm
GeV})$ was assumed at $M_G$ and the dominant LFV contributions to
$\bm^2_\tl$ were generated by RG evolution from $M_G$ down to the
decoupling scale $M_T$. In such a case, finite contributions like
those in eqs.~(\ref{finite}, \ref{soft2}) also emerge at $M_T$, but
they are subleading with respect to the RG corrections, since $B_T$
is of the same order as $m_0$. Instead, in the present picture,
there is a hierarchy between the SSB parameter $B_T$ and the
remaining ones [see eqs.~(\ref{finite}, \ref{soft2})], $B^2_T \gg
(B_T g^2/16 \pi^2)^2  \sim m^2_0$. However, in both scenarios the
flavor structure of $\bm^2_\tl$ is proportional to $\bY^\dagger_T
\bY_T$ and  can be written by using eq. (\ref{T-mass}) in terms of
the neutrino parameters (the terms $\propto g^2 \bY^\dagger_T\bY_T$
are generically the leading ones): 
\be (\bm^2_\tl)_{ij} \propto B_T^2 (\bY^\dagger_T \bY_T)_{ij} \sim
B_T^2 \left(\frac{M_T}{\la \langle H_2\rangle^2}\right)^{\!\!2}\!\!
\left[\bU (\bm^{D }_\nu)^2 \bU^\dagger\right]_{ij} . \ee
Consequently, the relative size of LFV in the different leptonic
families  can be univocally predicted as: 
\be
\label{predi}
\!\!\frac{ (\bm^{2 }_{\tilde{L}})_{\tau \mu}}
  {(\bm^{2 }_{\tilde{L}})_{\mu e} } \approx
\rho \frac{s_{23} c_{23}}{s_{12} c_{12} c_{23}}
\sim 40 ~ ,~
\frac{ (\bm^{2 }_{\tilde{L}})_{\tau e}}
  {(\bm^{2 }_{\tilde{L}})_{\mu e} } \approx -
\frac{s_{23}}{c_{23}} \sim - 1 ,
\ee
where $\rho= (m_3/m_2)^2$, $\theta_{12}$ and $\theta_{23}$ are
lepton mixing angles and $\theta_{13}=0$ is taken (the notation
$c_{ij}=\cos\theta_{ij}, \ldots$ is used). A hierarchical neutrino
mass spectrum is considered and the best-fit values for the
parameters~\cite{fit} are used. By taking the present upper limit on
$\sin\theta_{13}=0.2$, the above ratios become $3$ and $0.8$,
respectively, while varying  the other neutrino parameters within
their experimental range affect these ratios by less than $10\%$
(see also ~\cite{af2}). The above relations imply that also the
branching ratios (BR) of LFV processes such as the decays $\ell_i
\to \ell_j \gamma$ can be predicted
\be\label{brs} \frac{{\rm BR}(\tau \to \mu \gamma)} { {\rm BR}(\mu
\to e \gamma)} \sim 300 , ~~~~ \frac{ {\rm BR}(\tau \to e
\gamma)}{{\rm BR}(\mu \to e \gamma)} \sim 10^{-1} .
 \ee
Other LFV processes and related correlations~\cite{br} will be
considered in~\cite{af2}. (Connections between neutrino parameters
and other observables can arise also in different scenarios, see
{\it e.g.} \cite{others}). Without loss of generality we take $B_T$
to be real since its phase has not physical effect. However, a
different approach was considered in~\cite{cmrv} where a complex
$B_T$ could generate sizable electric dipole moments for quarks and
leptons since there was a relative phase between the trilinear
couplings $\bA_{e,d,u}$ shown in eq.~(\ref{finite}) and the gaugino
masses. Moreover, the soft term $B_T$ could play a significant role
in generating the baryon asymmetry of the Universe in the context of
resonant leptogenesis~\cite{lepto}.

We shall now discuss the phenomenological viability of our scenario.
Throughout our discussion we shall take $M_T > 10^{7}~{\rm GeV}$ so
that the gauge coupling constants remain perturbative up to $M_G$.
Our approach follows a bottom-up perspective where, for a given
ratio $M_T/\la$ and $\tan\beta$, $\bY_T$ is determined at $M_T$
according to the matching expressed by eq.~(\ref{T-mass}) using the
low-energy neutrino parameters. The Yukawa matrices $\bY_e, \bY_u,
\bY_d$ are determined by the related charged fermion masses, modulo
$\tan\beta$. Although the $\mu$-parameter is not predicted by the
underlying theory, it is nevertheless determined together with
$\tan\beta$ by correct EWSB conditions. Therefore, we end up with
only three free parameters, $B_T, M_T$ and $\la$.
%
\begin{figure}
\begin{center}
\begin{tabular}{c}
\includegraphics[width=13cm]{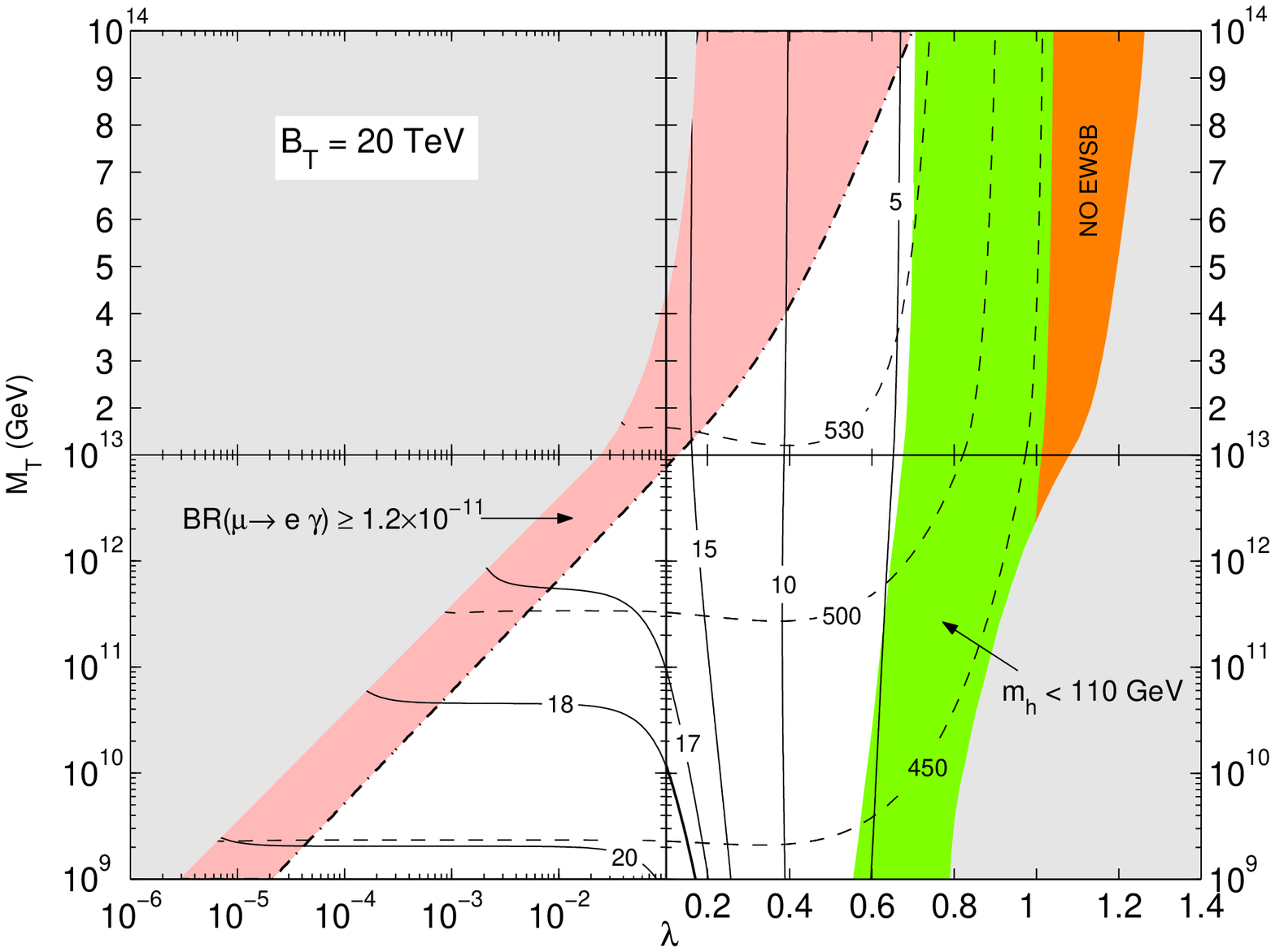}\\
\includegraphics[width=13cm]{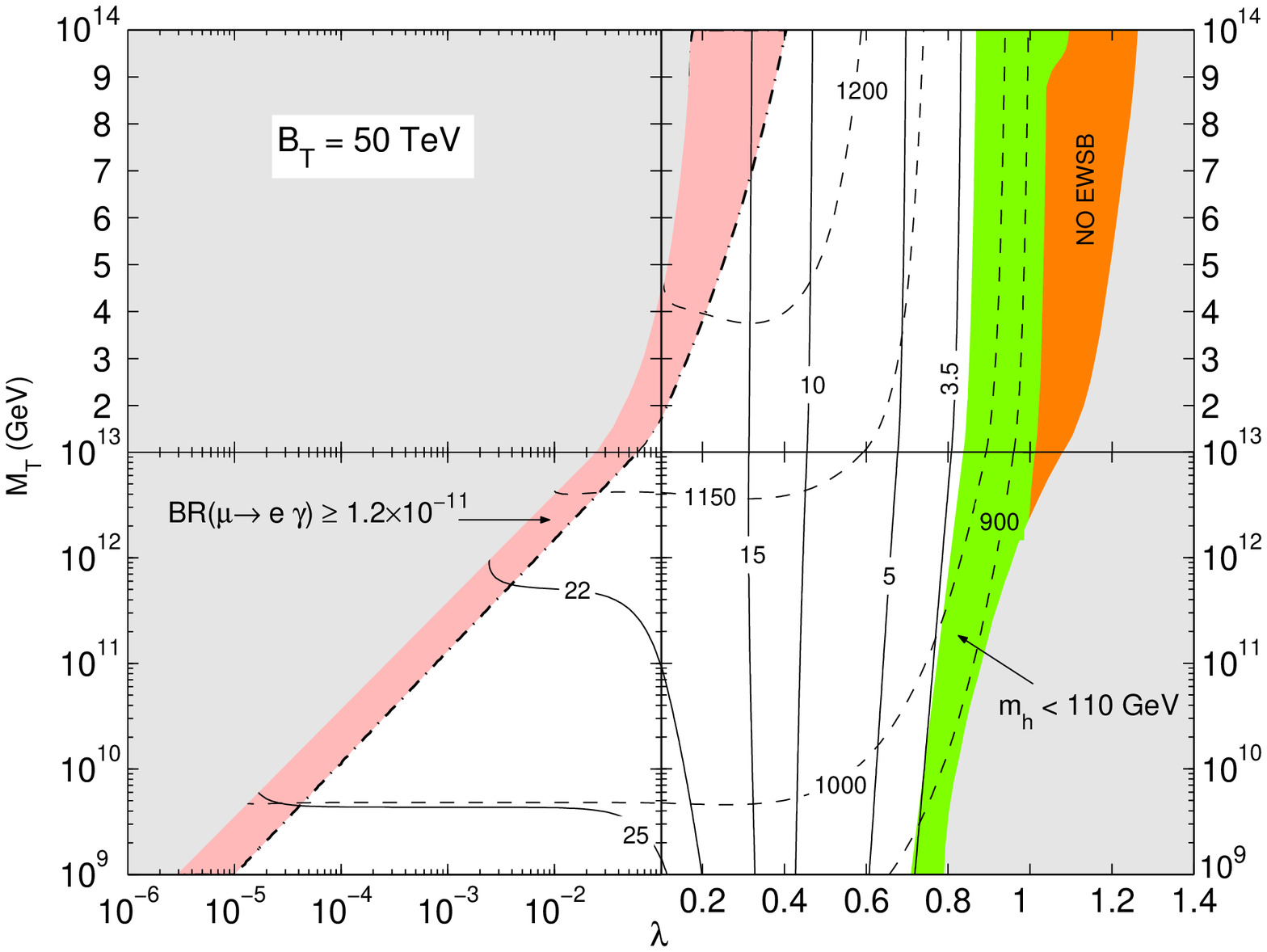}
\end{tabular}
\caption{\it The $(\la, M_T)$ parameter space constrained by the
perturbativity requirement (light-grey), correct EWSB  from one-loop
corrected Higgs potential, lower bound on the lightest Higgs boson
mass $m_h$ and the upper bound on $BR(\mu\to e \ga)$, for $B_T=20
(50)~{\rm TeV}$ in the upper (lower) panel.  We also display the
isocontours of $\tan\beta$ (solid) and $\mu$ (dashed). We have taken
the top pole mass $m_t= 174~{\rm GeV}.$ \vspace*{0.cm} } \label{f1}
\end{center}
\end{figure}
In Fig.~\ref{f1} we show the $(\la, M_T)$ parameter space  allowed
by the perturbativity and EWSB requirements, the experimental lower
bound on the lightest Higgs mass\footnote{We include the low-energy
radiative corrections to the Higgs masses by linking our code to
{\tt FeynHiggs} \cite{FH}.} $m_h$ and the upper bound on $BR(\mu \to
e \ga)$, for $B_T = 20\, (50)~{\rm TeV}$ in the upper (lower) panel.
First notice the light-grey regions excluded by the perturbativity
requirement which are independent of $B_T$. For each value of $M_T$
there is a minimum value of $\la$, which scales as $\sim 2\times
10^{-4}(M_T/10^{11}~{\rm GeV})$, below which the couplings $\bY_T$
reach the Landau pole below $M_G$. Similarly, there is a maximum
value of $\la$ beyond which $\la$ itself blows up below $M_G$. The
EWSB constraint excludes a region for $\la \sim 1 - 1.2$ and $M_T
\gsim 10^{12}~{\rm GeV}$ (independently of $B_T$) which is limited
by the least achievable value of $\tan\beta$, $\tan\beta\sim 2.5$.
As for the $\mu$-parameter (dashed lines), it slightly increases
with increasing $M_T$ due to the large RG factor which affects
$m^2_{H_2}(\mu_{SUSY})$ in the minimization condition, $
\mu^2(\mu_{SUSY}) \approx - m^2_{H_2}(\mu_{SUSY})$, covering the
range $\mu \sim 450 - 550 \,(1000 - 1200)~{\rm GeV}$ for $B_T
=20\,(50)~{\rm TeV}$. We observe that $\la < 0.6 \,(0.7)$ for $B_T
=20\, (50)~{\rm TeV}$ is required by the constraint $m_h> 110~{\rm
GeV}$. The related contour lies on the correspondent minimum
 value of $\tan\beta \sim 5 \,(3.5)$  for $B_T =20\,(50)~{\rm
TeV}$. When $B_T =50  ~{\rm TeV}$, the sparticle spectrum is heavier
and thus the radiative corrections $\sim \log
(\frac{\mu_{SUSY}}{m_t})$ to $m_h$ are larger and in the tree-level
contribution $\sim M_Z |\cos 2\beta|$ smaller $\tan\beta$ can be
tolerated.

The present bound on $BR(\mu \to e \ga)$ provides a lower bound on
$\la$ for each value of $M_T$. This stems from the fact that the LFV
entries $(\bm^2_\tl)_{ij}$ scale as $(M_T/\la)^2$
[eq.~(\ref{soft2})]. Consequently, the allowed $\la$-range is wider
for lower values of $M_T$ and, comparing the two panels, the whole
parameter space is larger for $B_T = 50~{\rm TeV}$. In the allowed
regions, the lightest MSSM sparticle is typically a charged slepton
with mass around $100-200\,(300-450) ~{\rm GeV}$ for
$B_T=20\,(50)$~TeV, although for small $\tan\beta$ there could be a
mass degeneracy with the lightest neutralino. However, either the
lightest slepton or neutralino would decay into the gravitino which
is most likely the lightest supersymmetric particle in our
framework. Finally, we have checked that values of $B_T < 10~{\rm
TeV}$ are phenomenologically unacceptable.

\begin{figure}
\begin{center}
\includegraphics[width=12cm]{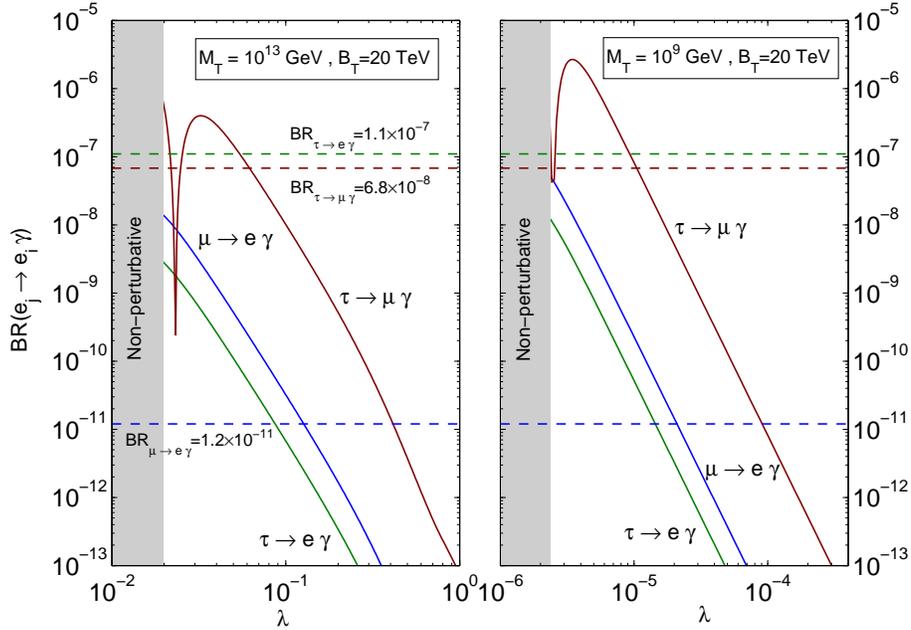}
\end{center}
\vglue -0.5cm \caption{\it The $BR$s of the lepton radiative decays
are shown as a function of $\la$ for $B_T=20~{\rm TeV}$ and for
$M_T=10^{13} (10^{9})~{\rm GeV}$ in the left (right) panel. The
horizontal lines indicate the present bound on each $BR$
\cite{exp}.} \label{f2} \vglue -0.6cm
\end{figure}

In Fig.~\ref{f2} we display the branching ratios $BR(\ell_j \to
\ell_i \gamma)$ as a  function of $\la$ for $B_T= 20~{\rm TeV}$
 and $M_T= 10^{13}\,(10^{9})~{\rm GeV}$
in the left (right)  panel. The behaviour of these branching ratios
is in remarkable agreement with the estimates of eq.~(\ref{brs}).
Hence, the relative size of LFV does not depend on the detail of the
model, such as the values of $\la$, $B_T$ or $M_T$. This feature is
not present for a very narrow range of  $\la$ where $BR(\tau\to \mu
\ga)$ is strongly suppressed due to a conspiracy of the various
contributions in $(\bm^2_\tl)_{\tau \mu}$ which mutually cancel out
[see eq.(\ref{soft2})].

Before concluding, we would like to briefly mention that the
tree-level exchange of the $T,\bar{T}$ states also generates the
$L$-violating SSB operator $\lambda\bY_T B_T(\tl H_2)^2/M_T$ which
induces a sneutrino/anti-sneutrino mass splitting $\Delta
\bm^2_{\tilde{\nu}}=B_T\bm_\nu$ at the EW scale. Since $B_T$ is much
larger than the EW scale, we are led to think that this could render
interesting effects for the phenomenology of sneutrino
oscillations~\cite{GH}.

In conclusion, we have suggested a unified picture of the
supersymmetric type-II seesaw where the triplets, besides being
responsible for neutrino mass generation, communicate SUSY breaking
to the observable sector through gauge and Yukawa interactions. We
have performed a phenomenological analysis of the allowed parameter
space emphasizing the role of LFV processes in testing our
framework.

\vspace{0.5cm}

{\bf Acknowledgments:} We thank A.~Brignole for useful comments. The
work of F.R.J. is supported by {\em Funda\c{c}{\~a}o para a
Ci{\^e}ncia e a Tecnologia} (FCT, Portugal) under the grant
\mbox{SFRH/BPD/14473/2003},  INFN and PRIN Fisica Astroparticellare
(MIUR). The work of A.~R.~ is partially supported by the project EU
MRTN-CT-2004-503369.

\vglue -0.4cm

\end{document}